\documentclass[%
reprint,
 amsmath,amssymb,
prb,
nolongbibliography,
]{revtex4-2}

\usepackage{graphicx}
\usepackage{dcolumn}
\usepackage{bm}
\usepackage{hyperref}

\usepackage{float}
\usepackage{tikz}
\usepackage{physics}
\usepackage{subcaption} 
\usepackage{xcolor}     
\usepackage{booktabs}   
\usepackage{tabu}       

\newcolumntype{L}{>{$}l<{$}}
\newcolumntype{C}{>{$}c<{$}}
\newcolumntype{R}{>{$}r<{$}}

\newcommand{\dfthalf}{DFT-$\frac{1}{2}$ }
\newcommand{\gammapoint}{$\Gamma$-point }

\begin{document}


\title{The decoupled \dfthalf method for defect excitation energies}

\author{Joshua Claes}
\author{Bart Partoens}%
\affiliation{%
CMT, Physics Department, University of Antwerp. Groenenenborgerlaan 171, 2020 Belgium
}%

\author{Dirk Lamoen}
\affiliation{
EMAT, Physics Department, University of Antwerp. Groenenenborgerlaan 171, 2020 Belgium
}%

\date{September 7, 2023}

\begin{abstract}
The \dfthalf method is a band gap correction with GW precision at a DFT computational cost. The method was also extended to correct the gap between defect levels, allowing for the calculation of optical transitions. However, this method fails when the atomic character of the occupied and unoccupied defect levels are similar as we illustrate by two examples, the tetrahedral hydrogen interstitial and the negatively charged vacancy in diamond.  We solve this problem by decoupling the effect of the occupied and unoccupied defect levels and call this the decoupled \dfthalf method for defects.

\end{abstract}

\maketitle


\section{\label{sec:Intro}Introduction}
Creating point defects in solids allows one to manipulate the properties of that solid for example by doping the material to obtain a P or N type semiconductor. The defects themselves can also be interesting subsystems 
in the context of quantum technologies, where defects like the NV center \cite{NV_colour_center_in_diamond_DOHERTY, gali2019} have the potential to be used as a quantum sensor or as qubits, the building block for the quantum computer \cite{Color_centers_in_diamond_THIERING20201,Quantum_computer_based_on_color_centers_in_diamond}.  The primary approach to simulate these defects is density functional theory (DFT). However, DFT is known to underestimate the band gap of semiconductors for the LDA or GGA exchange correlation functionals. This is due to the local nature of these approximate exchange correlation functionals, which neglects the exchange correlation discontinuity at the band gap \cite{XC_discont_PhysRevLett.51.1884, DFT_of_the_bandgap_PhysRevB.32.3883, Mao_2022}. When the band gap is a property of interest, as is the case with defects in solids, one has to rely
on more advanced methods such as meta-GGA functionals like SCAN \cite{SCAN_PhysRevLett.115.036402} or hybrid functionals like HSE06 \cite{HSE03,HSE06} or GW.

In 2008 Ferreira et al. \cite{Ferreira2008} introduced a new method, the \dfthalf method, which rectifies the lack of self-energy of the band gap. This is achieved by adding a self-energy potential to the pseudopotential in such a way that the self-energy is added to the band gap. The \dfthalf method was later expanded upon by Lucatto et al. \cite{dfthalf_NVmin} to work for defect levels. However, when these defect levels have a similar orbital character, the \dfthalf method will fail. Prior to this the \dfthalf method has also been used to calculate the formation and transition energy of an interstitial/substitutional Mn defect and a self-interstitial in silicon \cite{ChargeTransMn, Efrom_Si_dft12}.

In this work we show that defect levels with similar orbital characters will cause the self-energy potential to be approximately zero, negating the \dfthalf correction. This is first illustrated on the tetrahedral hydrogen interstitial in diamond, an example chosen such that the problem is maximal. 
The goal of this work is to show why the self-energy potential is zero in these cases and to introduce a new \dfthalf technique to solve this problem: \emph{The decoupled \dfthalf method for defects}. The hydrogen interstitial is first revisited using the new method. It is then shown that the band structure of the defect system can be reconstructed and it is compared to the band structure obtained from calculation with the HSE06 functional,
a widely used functional in high quality defect calculations \cite{GaliDeak_FormationNV,Pershin2021,DEAK201835, RevModPhys.86.253}.
In addition to the hydrogen interstitial the negatively charged vacancy in diamond will also be studied with the decoupled \dfthalf method. The negatively charged vacancy was chosen because vacancies or vacancy related defects are a class of defects that likely require the decoupled \dfthalf method to calculate defect gaps. The reason for this is that the creation of a vacancy leaves behind dangling bonds from atoms of the host material, usually in some sort of symmetric configuration. This makes it likely that defect levels with similar orbital character appear. 

\subsection{The \dfthalf method}
We now give a brief overview of the \dfthalf method, where we focus on those parts that are important to introduce the decoupled \dfthalf method. More details can be found in the original papers by Ferreira et al. \cite{Ferreira2008,Ferreira2011} or in the recent review of Mao et al. \cite{Mao_2022}. The \dfthalf method starts with Janak's theorem and the assumption that the Kohn Sham (KS) eigenvalue $\varepsilon_\alpha$ of orbital $\alpha$ is linearly dependent on the occupation of the orbital $f_\alpha$, the validity of this assumption was verified in Ref. \cite{Leite_1971}. With this, one can derive that the band gap of a semiconductor, which is the difference between the ionization energy $I$ and the electron affinity $A$, is equal to the difference between the eigenvalue of the half occupied conduction band minimum (CBM) and valence band maximum (VBM) or
\begin{align}
    \text{Band gap}	&= I - A\\
			        &= \left( E_{tot}^{N-1} - E_{tot}^{N} \right) -  \left( E_{tot}^{N} - E_{tot}^{N+1} \right)\\
			        &= \varepsilon_c(f_c=1/2) - \varepsilon_v(f_v=-1/2) \label{eq:BGelectransfer} 
\end{align}
where the sub-indices $c$ and $v$ denote the CBM and VBM, respectively, and $f_\alpha=0$ indicates that orbital $\alpha$ has the same occupation as in the ground state and $f_\alpha=\pm 1/2$ means that half an electron was added or subtracted from the ground state occupation. From Janak's theorem a new quantity can be derived: the \emph{self-energy} $S_\alpha$ which is defined by 
\begin{equation} 
\pdv{\varepsilon_\alpha}{f_\alpha} = 2 S_\alpha \label{eq:twoSalpha}.
\end{equation}
By integrating Eq. \eqref{eq:twoSalpha}, Eq. \eqref{eq:BGelectransfer} can be rewritten in terms of the KS-gap and the self-energy:
\begin{equation}
    \text{Band gap}	= \text{KS-gap} + S_c + S_v \label{eq:BGselfE}
\end{equation}
Equation \eqref{eq:BGelectransfer} gives the impression that one could calculate the band gap with DFT by placing half an electron from the valence band in the conduction band. However, since the KS-eigenstates are Bloch states, which are delocalized, the self-energy of these states will be zero. Thus, according to \eqref{eq:BGselfE} this approach gives no correction to the KS-gap or in other words, Bloch states do not accurately describe the localized holes \cite{Ferreira2011}. Instead of changing the occupation within a calculation, a potential $V_s$ is added to the pseudopotential as if the occupation was changed or as if the self-energy was added. This potential $V_s$ is called the \emph{self-energy potential} and the self-energy can be seen as a quantum mechanical average over this potential. The self-energy potential can be calculated as follows:
\begin{equation}
V_s = V_{KS}(f_\alpha=0, \mathbf{r}) - V_{KS}\left(f_\alpha = -1/2, \mathbf{r} \right) \label{eq:VS_KS}
\end{equation}
with $V_{KS}$ the KS-potentials where the dependency on the electron density is not written explicitly and only the occupation of orbital $\alpha$ is considered as all other occupations remain the same.
The KS-potentials of Eq. \eqref{eq:VS_KS} are usually calculated for a single isolated atom using an all electron code.
The Coulomb-like tail of the self-energy and the periodic boundary condition imposed in a DFT calculation will lead to a divergence. This divergence can be removed by defining a new self-energy potential $\tilde{V}_s(\mathbf{r}) = \Theta(\mathbf{r}) V_s(\mathbf{r})$, where $\Theta(\mathbf{r})$ is a trimming function defined as 
\begin{align}
 \Theta(\mathbf{r})= 
    \begin{cases}
    \left( 1-\left(\frac{r}{r_c}\right)^n\right)^3 &r \leq r_c \\
    0 &r > r_c
    \end{cases} \label{eq:TrimmingFunction}
\end{align}
The trimming function introduces two new parameters $n$ and $r_c$ to the self-energy potential. The former is usually set to 8 as this gives a good balance between the cutoff sharpness and the potential smoothness \cite{Mao_2022}. The parameter $r_c$ is called the cutoff radius and should be determined by extremizing the band gap \footnote{In practise this usually means maximizing the band gap as DFT underestimates the band gap}. This means that in order to calculate the band gap using \dfthalf one should sweep over multiple DFT calculations with self-energy potentials at different cutoff radii.

\section{Method}
\subsection{The conventional \dfthalf method for defects}

\begin{figure}[h]
    \centering
    \includegraphics[width=0.4\textwidth,trim={3cm 2.5cm 3cm 3cm},clip]{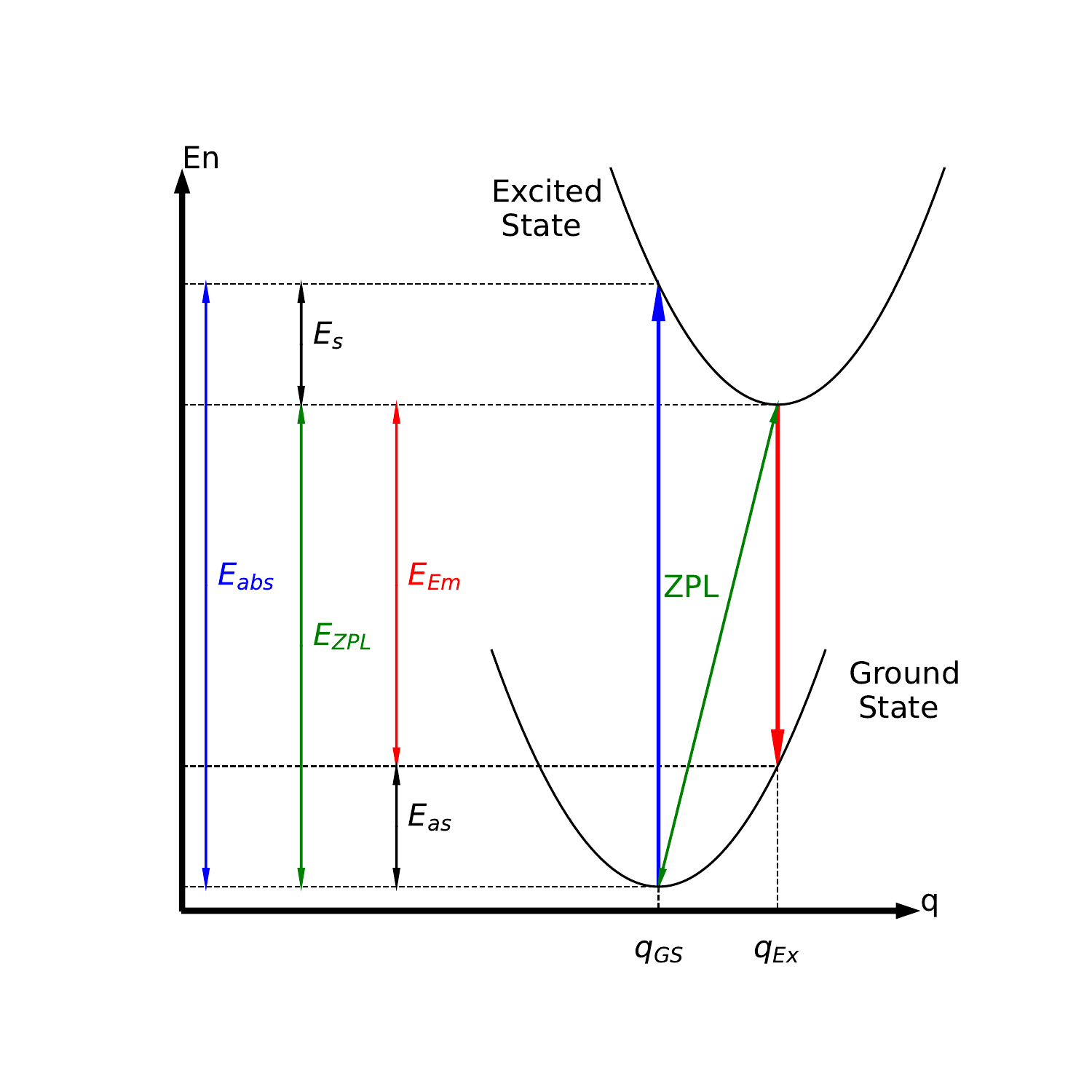}
    \caption{A schematic overview of the energies involved in excitation and deexcitation of a defect. On the abscissas we have the configuration space $q$ with the ground state configuration of the defect in its ground and first excited state denoted as $q_{GS}$ and $q_{Ex}$, respectively. The blue and red lines are the absorption $E_{abs}$ and emission energy $E_{em}$, respectively, and can be calculated using the \dfthalf method.}
    \label{fig:schema_ZPL}
\end{figure}

Since the \dfthalf method only uses DFT calculations the method has DFT computational scaling. This makes it an attractive method for defect calculations where a large supercell and a correct band gap are required. In Ref. \cite{dfthalf_NVmin} Lucatto et al. introduced the \dfthalf method for defect excitations. In this work, the \dfthalf method is used to calculate the gap between an occupied and unoccupied defect level in the band gap. This result can then be used to calculate the vertical transition energy between the ground and excited state of a defect, i.e. the absorption $E_{abs}$ and emission energy $E_{em}$, as is also illustrated in Fig. \ref{fig:schema_ZPL}. By calculating the Stokes or anti-Stokes shift, which must be done by plain DFT as this is a difference between total energies of two different structures, the zero phonon line (ZPL) can be obtained. The ZPL is an important and identifying property for color centers. Lucatto et al. \cite{dfthalf_NVmin} demonstrate this procedure for the $NV^-$ center in diamond and find a ZPL of $1.84$ eV close to the experimentally observed $1.95$ eV \cite{Davies_1976_NVmin}.

The \dfthalf method proposed by Lucatto et al., which we will call the conventional \dfthalf method for defects from now on, starts by dividing the atoms in the supercell into two groups: the defect atoms which are responsible for the defect levels and the bulk atoms which have a negligible contribution to the defect levels and which are responsible for the valence and conduction bands. 
In order to use the \dfthalf corrected band gap in a defect calculation, all bulk atoms should use the pseudopotential determined by the \dfthalf method on the pristine host material of the defect.
The group of defect atoms still has an unaltered pseudopotential allowing for an additional \dfthalf correction for the defect levels to be applied by adding a self-energy potential to these atoms. The general idea of the conventional method is the same, i.e. half an electron should be moved from the occupied to the unoccupied defect level in order to add self-energy. 
Instead of removing half the electron from the orbital with the largest contribution to the occupied defect level, a smaller fraction is removed from every orbital of every atom contributing to the defect level. 
The fraction removed from orbital $\phi$ of atom $X$ is called
\begin{equation}
    \xi_{X_\phi} =  \frac{1}{2} \, \text{char}_{X_\phi}[\psi_\alpha(\Gamma)]\label{eq:Xi_elecfrac}
\end{equation}
with char$_{X_\phi}[\psi_\alpha(k)]$ the projection of KS-state $\psi_\alpha$ of orbital $\alpha$ on to the atomic orbital $\phi$ of atoms $X$. Since exactly half an electron should be removed, $\xi_{X_\phi}$ should be normalized such that
\begin{equation}
    \sum_{X_\phi} \xi_{X_\phi} = \frac{1}{2} \label{eq:Xi_norm}
\end{equation}
For the unoccupied level the fraction of electron which will be added to each orbital of each defect atom $\zeta_{X_\phi}$ is calculated in a similar fashion. The self-energy potential for each orbital of each atom is then calculated as 
\begin{equation}
    V^{X_\phi}_S = V_X^{KS}(f_0-\zeta_{X_\phi}) -V_X^{KS}(f_0-\xi_{X_\phi}) \label{eq:VS_defect}
\end{equation}
where only the occupation of the orbital $\phi$ is written as an input for $V_X^{KS}$ as all other inputs are the same and with $f_0$ the ground state occupation of that atom. As is the case with bulk \dfthalf, the self-energy potentials need to be multiplied by a trimming function $\Theta_{X_\phi}$. In the most general case, each orbital of each atom has its own trimming function and cutoff radius. However, we will assume that each orbital of the same atom has the same trimming function. Thus, the total self-energy potential of each atom is given by
\begin{equation}
    V_S^X = \Theta_X \sum_\phi V^{X_\phi}_S
\end{equation}
where $\Theta_X$ is given by Eq. \eqref{eq:TrimmingFunction}. To find the cutoff of each atom, the gap between the defect levels needs to be extremized consecutively. 

The problem that can appear in the conventional \dfthalf method is best illustrated on the tetrahedral hydrogen interstitial in diamond.

\subsection{\label{sec:Comp_det}Computational details}
The DFT calculations were performed in the local density approximation (LDA) exchange correlation potential and PAW \cite{vasppaw} as implemented by the Vienna \textit{ab initio} simulation package (VASP) \cite{vaspcode1,vaspcode2,vaspcode3} taking (collinear) spin polarization in to account. We chose LDA because Janak's theorem is exact for the LDA exchange correlation functional \cite{Ferreira2011}. 
We calculate the lattice parameter of the conventional unit cell of diamond using a Birch-Murnagahan fit \cite{BMeq} with a cutoff energy of $520$ eV and $8\times8\times8$ $k$-point grid in the Monkhorst-Pack scheme \cite{Monkhorst_Pack_scheme}. This gives us a distance between carbon atoms and a lattice parameter of $1.53\enspace \text{\AA}$  and $3.54\enspace \text{\AA}  $, respectively,  which is in good agreement with the experimental values \cite{semiconductor_properties}. The defect supercells were created from a $4\times4\times4$ conventional diamond supercell with 512 carbon atoms. The integration over the Brillouin zone was done using only the \gammapoint with an energy cutoff of $520$ eV. Since the \dfthalf method does not produce a correct total energy, all relaxations were done using LDA. The relaxations were stopped when all forces are below $0.001\enspace \text{eV}/\text{\AA}$.

The KS-potentials used for calculating the self-energy potential in Eq. \eqref{eq:VS_KS} in the \dfthalf calculation were generated using a modified version of the \textsc{atom} code \cite{Siesta_package,Ferreira2008}.
The \dfthalf band gap for diamond was calculated by stripping $1/4$th of both the $s$ and $p$ orbital as suggested by Ferreira et al. \cite{Ferreira2008}. This results in a band gap of $5.73$ eV and a cutoff parameter of $2.3\enspace a_0$, which is in line with the results obtained in Ref. \cite{limitsdfthalf,Xue_shdfthalf}.

\subsection{The hydrogen interstitial in diamond with the conventional method}
In this section the electronic structure of the interstitial hydrogen defect in diamond is calculated using the conventional \dfthalf method. The nature of this defect makes it likely to have at least two defect levels with the same character, namely the $H_{s,\uparrow}$ and $H_{s,\downarrow}$ orbitals localized around the hydrogen atom. It is then demonstrated that the conventional method does not improve the DFT defect gap.

Interstitial hydrogen in diamond can either be found in a negative, positive or neutral charged state. We will focus on the neutral charged state, because in this charge state the defect level has one occupied and one empty defect level which is required by the conventional \dfthalf method. 
In the neutral state, interstitial hydrogen in diamond has 3 stable configurations. Going from lowest to highest energy these states are named BC (bond center), T (tetrahedral) and H (hexagonal) \cite{Hint_dia_Goss, Lyons_2016}.
\begin{figure*}
    \centering
    \includegraphics[width=\textwidth]{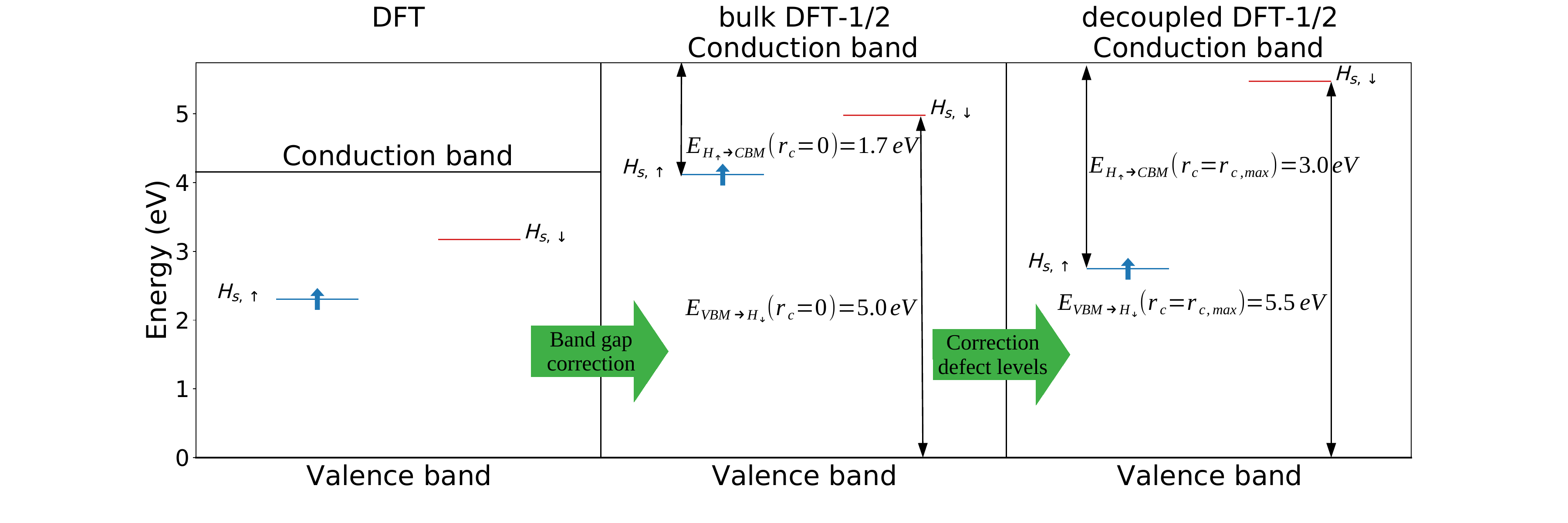}
    \caption{The band structure at the \gammapoint for the tetrahedral interstitial hydrogen in diamond, using LDA with an unmodified pseudopotential (left), LDA with the \dfthalf corrected pseudopotentials for bulk carbon atoms (middle) and the \dfthalf correction for bulk and defect atoms (right). The correction for the defect levels was obtained from the decoupled \dfthalf method. Because the decoupled method uses a different calculation for each defect level, the defect levels of the bulk \dfthalf band structure were shifted to their correct position with respect to the VBM and CBM as described in the text.}
    \label{fig:H_T_schema_BS}
\end{figure*}
Because the BC hydrogen defect was not suited for the \dfthalf method \footnote{As a first attempt the conventional \dfthalf method was tried on the BC configuration of the H interstitial i.e the lowest energy state. However, placing the hydrogen atom so close to the carbon atoms results in defect level with a non-negligible contribution from the $p$-orbitals of the carbon atoms that form the bond and the $p$-orbital of hydrogen. This means that these carbon atoms should be considered as defect atoms. 
The contribution of the $H_p$ orbital is problematic for the \dfthalf method in general, because the method requires that a fraction of the empty $p$-orbital of the hydrogen atom in its ground state is removed. This problem will not be solved in this paper but could be the subject of a future study.}, 
the lowest meta-stable hydrogen interstitial, tetrahedral hydrogen, in diamond was studied. In this structure the hydrogen is located in one of the cavities of the diamond supercell. If this cavity and the hydrogen atom are placed along the $[111]$ direction, then the hydrogen atom will be closer to one of the carbon atoms along this direction. If instead the hydrogen atom has the same distance to both carbon atoms along the $[111]$ direction, then the defect is in the hexagonal state \cite{Hint_dia_Goss}.
After relaxation the energy of the hydrogen interstitial is $1.09$ eV higher than the energy of the BC configuration. This energy difference between the ground and meta-stable state is similar to that found in \cite{Hint_dia_Goss, UPADHYAY2014257} and deviates somewhat from \cite{Lyons_2016}, although this is likely because they use HSE06 instead of LDA. 

\begin{figure}
    \centering
    \includegraphics[width=0.30\textwidth]{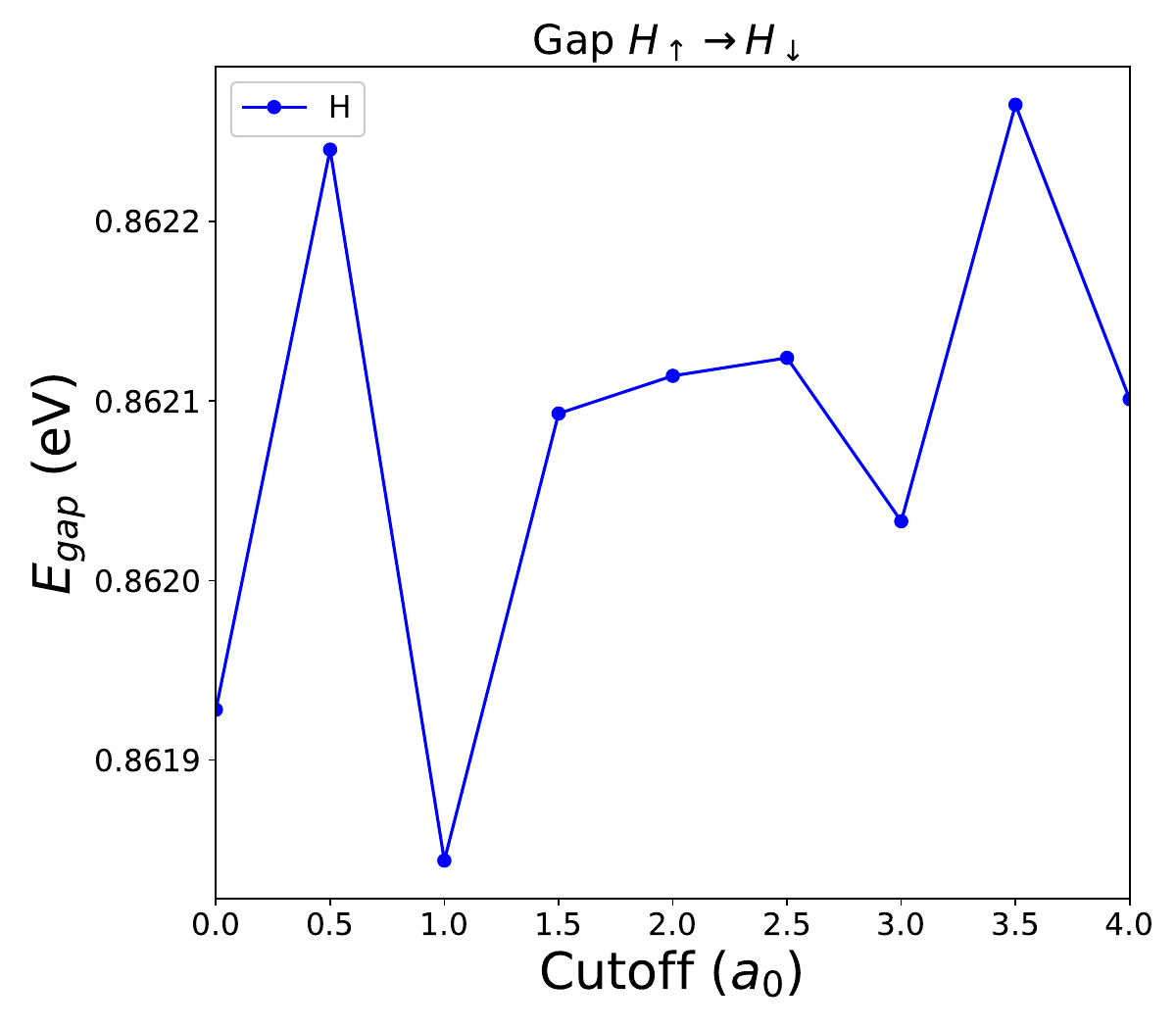}
    \caption{The defect gap maximization for the interstitial tetrahedral hydrogen defect in diamond using the conventional \dfthalf method.}
    \label{fig:H_T_conv}
\end{figure}

In Fig. \ref{fig:H_T_schema_BS} the \gammapoint band structure of the tetrahedral hydrogen defect is shown. In the case of tetrahedral hydrogen the only significant contribution to the defect levels of Fig. \ref{fig:H_T_schema_BS} comes from the $s$-orbital of the hydrogen atom, meaning that hydrogen is our only defect atom and $\xi_{H_s}=\zeta_{H_s}=0.50$. With this the self-energy was determined with formula \eqref{eq:VS_defect} and the gap between the defect levels was determined using the conventional \dfthalf method. The gap between the defect levels seems to be unaffected by the cutoff parameter, as can be seen in Fig. \ref{fig:H_T_conv}, the only difference is the run-to-run variance between the DFT runs. The reason for this failure of the conventional method will be explored in the next section.

\subsection{The decoupled \dfthalf method}
\begin{figure}
    \centering
    \includegraphics[width=0.45\textwidth]{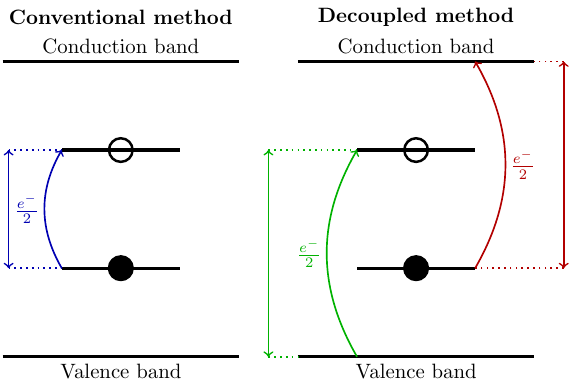}
	\caption{A schematic of the KS-band structure with one occupied and unoccupied defect level depicted by the filled and unfilled circle, respectively. On the left the electron transfer of the conventional method is shown by the blue arrow. On the right the gaps $E_{VBM \rightarrow unocc}$ and $E_{occ \rightarrow CBM}$ as well as the corresponding electron transfers of the decoupled method are drawn by the green and red arrow, respectively.}\label{fig:schema_dec_method}
\end{figure}

The self-energy potential used in the \dfthalf method is spherically symmetric,  meaning that only the $s$, $p$ or $d$ orbital character is looked at and the method does not differentiate between $p_x$, $p_y$ and $p_z$ for example. 
For some defects this results in a situation where $\xi_{X_ \phi} \approx \zeta_{X_ \phi}$, 
as was the case with the tetrahedral hydrogen interstitial. 
In these cases it follows that
\begin{equation}
    V_X^{KS}(f_0-\zeta_{X_\phi}) \approx V_X^{KS}(f_0-\xi_{X_\phi}) \label{eq:VxiapproxVzeta}
\end{equation}
By using Eq. \eqref{eq:VxiapproxVzeta} in \eqref{eq:VS_defect} we see that both contributions to the self-energy potential for $X_\phi$ cancel:
\begin{equation}
    V^{X_\phi}_S = V_X^{KS}(f_0-\zeta_{X_\phi}) -V_X^{KS}(f_0-\xi_{X_\phi}) \approx 0
\end{equation}
which means that almost no self-energy is added when $\xi_{X_ \phi} \approx \zeta_{X_ \phi}$. This can be problematic when $\xi_{X_ \phi}$ and $\zeta_{X_ \phi}$ are relatively large and thus a large part of the \dfthalf correction should come from the self-energy potential of orbital $\phi$ of atom $X$. Because the self-energy potential is approximately zero, almost no correction is added to the defect gap. To put it simply the conventional method does not work for these cases because it removes and then adds the same electron fraction to the defect atoms.\\
\linebreak
Generally, this cancellation of the self-energy potential is the result of the approximation that the self-energy potential is spherical symmetric and thus this effect is unintended. To remove this cancellation of the self-energy the effect of $V_X^{KS}(f_0-\xi_{X_\phi})$ should be decoupled from that of $V_X^{KS}(f_0-\zeta_{X_\phi})$. This can be achieved by doing two separate calculations where either $V_X^{KS}(f_0-\xi_{X_\phi})$ or $V_X^{KS}(f_0-\zeta_{X_\phi})$ is added to the pseudopotential. Because $\xi_{X_\phi} \approx \zeta_{X_\phi}$ both the occupied and unoccupied level will move up or down together and the gap cannot be directly extremized as function of the cutoff. Instead, the defect gap should be extremized indirectly by extremizing $E_{VBM \rightarrow unocc}$: the gap between the VBM  and the unoccupied level and $E_{occ \rightarrow CBM}$: the gap between the occupied level and the CBM. 
In Ref. \cite{ChargeTransMn} the cutoff parameter of Mn is determined in a similar fashion, by following the defect level in the DOS as a function of $r_c$ with respect to the CBM.  
Both energy gaps are depicted in Fig. \ref{fig:schema_dec_method}, where the electron transfer of the conventional and decoupled method are illustrated. Because both the valence and conduction bands have already had a bulk \dfthalf correction and the effect of the defect atoms on these bands is negligible, both of these bands can be used as reference bands. 

These two gaps and the previously calculated \dfthalf band gap can formally be written as
\begin{align}
  E_{VBM \rightarrow unocc} &= \varepsilon_{unocc}^+ - \varepsilon_{VBM}^- \label{eq:vbmtounocc}\\  
  E_{occ \rightarrow CBM}   &= \varepsilon_{CBM}^+   - \varepsilon_{occ}^-\label{eq:occtocbm}\\
  E_{bandgap}               &= \varepsilon_{CBM}^+   - \varepsilon_{VBM}^- \label{eq:bg}
\end{align}
where $E$ is used for energy gaps, $\varepsilon$ for the KS-eigenvalues and the $+$ and $-$ super-indices are used to denote that these are the eigenvalues with half an electron added or subtracted, respectively. By using \eqref{eq:vbmtounocc}, \eqref{eq:occtocbm} and \eqref{eq:bg} the formula of the defect gap $E_{gap,def}$ can be written in terms of the other energy gaps as follows:
\begin{align}
    E_{gap, def}    =& \varepsilon_{unocc}^+ - \varepsilon_{occ}^- \\
                    =& \left(E_{VBM \rightarrow unocc} + \varepsilon_{VBM}^- \right) \nonumber \\
                    & -\left(\varepsilon_{CBM}^+ - E_{occ \rightarrow CBM}\right) \\
                    =& E_{VBM \rightarrow unocc} + E_{occ \rightarrow CBM} \nonumber \\
                    & - \left(\varepsilon_{CBM}^+  - \varepsilon_{VBM}^- \right) \\
                    =& E_{VBM \rightarrow unocc} + E_{occ \rightarrow CBM} - E_{bandgap} \label{eq:Egap_dec}
\end{align}
And thus with Eq. \eqref{eq:Egap_dec} the gap between defect levels $E_{gap, def}$ has successfully been rewritten in terms of the energy gaps $E_{VBM \rightarrow unocc}$ and $E_{occ \rightarrow CBM}$ and the defect gap can be calculated based on two decoupled calculations. Because there are now two gaps and two cutoff parameters $r_c$ to be determined the computational expense has doubled but the DFT scaling remains.

Since the defect gap in Eq. \eqref{eq:Egap_dec} depends on three quantities which are directly related to the band gap, any error made on this band gap by the bulk \dfthalf method will be carried over multiple times to the defect gap. We now determine the error of the defect gap as a result of the band gap error $\Delta$, where we assume that the band gap error is the only source of error in the calculation, 
such that the energy eigenvalue of the defect level is bound by the energy eigenvalue of the defect level when the gap between this level and the VBM is exact and the energy eigenvalue of the defect level when the gap between it and the CBM is exact. 
The experimental band gap can be written as follows
\begin{equation}
    E_{bandgap,exp} = E_{bandgap, DFT-1/2} + \Delta
\end{equation}
where the sign of $\Delta$ can be either positive or negative. In what follows $\Delta$ is assumed to be positive but a similar expression can be obtained when $\Delta$ is negative. 
When $\Delta=0$ the predicted \dfthalf gap and the actual gap are the same.
In cases where $\Delta \neq 0$, the \dfthalf predicted gaps can have the following values:
\begin{align}
    E_{VBM \rightarrow unocc}^{DFT-1/2} &\in [E_{VBM \rightarrow unocc}, E_{VBM \rightarrow unocc} + \Delta ] \label{eq:domain_EVBM_to_unocc} \\
    E_{occ \rightarrow CBM}^{DFT-1/2}   &\in [E_{occ \rightarrow CBM} , E_{occ \rightarrow CBM}  + \Delta] \label{eq:domain_Eocc_to_CBM}
\end{align}
where the super-index $DFT-1/2$ was added to denote the difference between the \dfthalf gap and the actual gap. 
Within this assumption the predicted values are either correct or at most a value $+\Delta$ off from the real values. 
This leaves 2 worst case scenarios where the error on $E_{gap,def}$ is maximal. The first being that both $E_{VBM \rightarrow unocc}$ and  $E_{occ \rightarrow CBM}$ are predicted correctly leaving an error of $-\Delta$ coming from the band gap term in \eqref{eq:Egap_dec}. The second happens when both $E_{VBM \rightarrow unocc}$ and  $E_{occ \rightarrow CBM}$ have an error of $\Delta$. One of these will be canceled by the error on the band gap while the other remains. These 2 worst case scenarios give the decoupled method an error margin of $\pm \Delta$ on $E_{gap,def}$ stemming from the band gap error.

\section{Results}

\subsection{The hydrogen interstitial in diamond with the decoupled method}
\begin{figure}[h]
    \centering
    \includegraphics[width=0.47\textwidth]{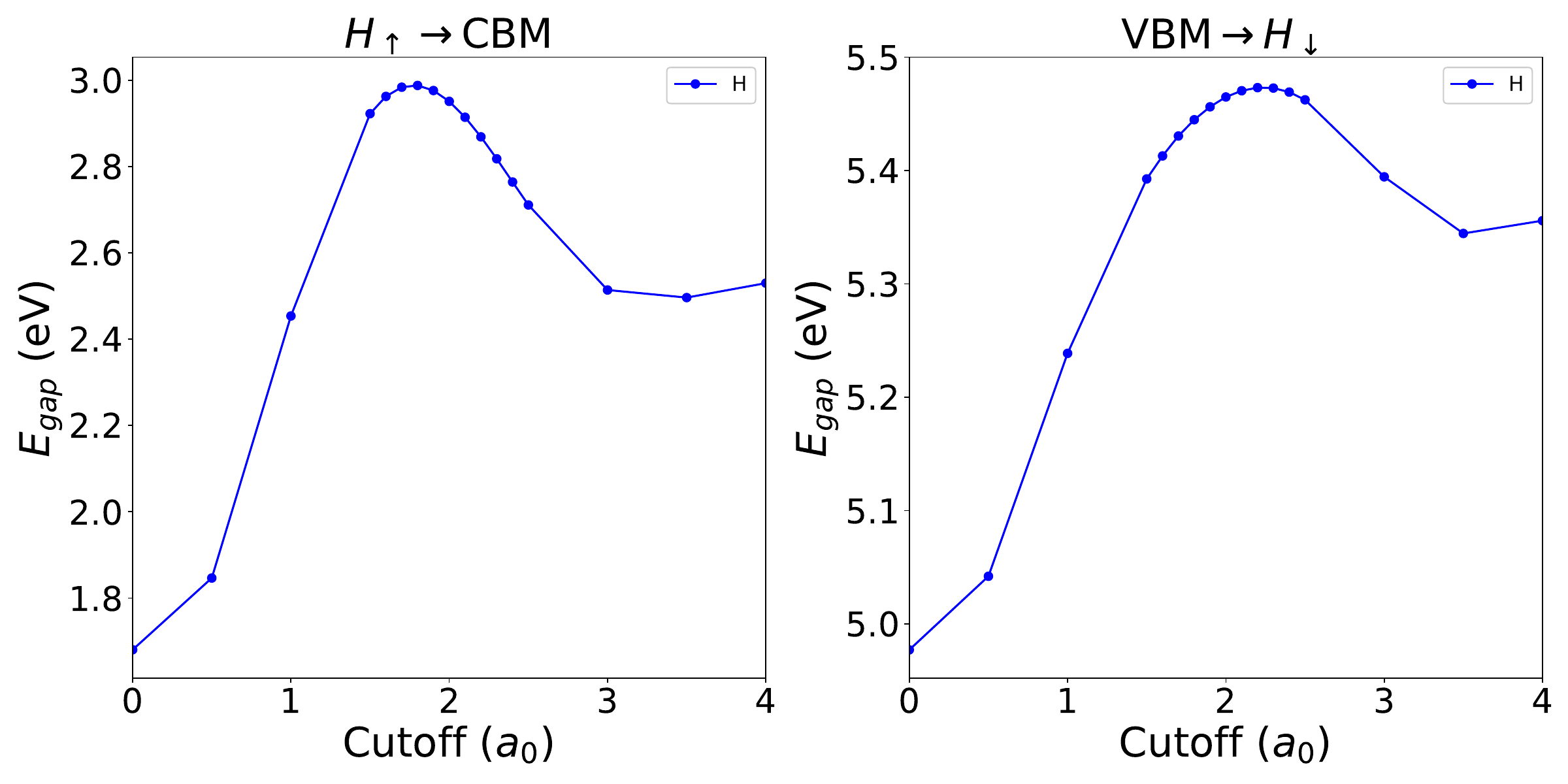}
    \caption{The maximization of the gap between occupied spin up level and the CBM (left) and the VBM and the unoccupied spin down level (right) for $H_{int,T}$ used for the decoupled \dfthalf method.}
    \label{fig:H_T_decoupled}
\end{figure}

Now that we have the tools to deal with cases where $\xi_{X_\phi}\approx \zeta_{X_\phi}$, we revisit the tetrahedral hydrogen interstitial in diamond. 
In Fig. \ref{fig:H_T_decoupled} the cutoff parameters and sizes of the gaps for $E_{VBM \rightarrow H_{s,\downarrow}}$ and  $E_{H_{s,\uparrow} \rightarrow CBM}$ are determined. With the maximum value of $E_{VBM \rightarrow H_{s,\downarrow}}$, $E_{H_{s,\uparrow} \rightarrow CBM}$ and Eq. \eqref{eq:Egap_dec} the gap between the $H_{s,\uparrow}$ and $H_{s,\downarrow}$ level was determined to be $2.7$ eV, about 3 times larger than the gap of $0.9$ eV when only the bulk \dfthalf correction is applied.

With the results of the decoupled \dfthalf method the band structure of the hydrogen interstitial was reconstructed starting from the bulk \dfthalf band structure. This bulk \dfthalf band structure is obtained from a DFT calculation where the bulk carbon atoms use the \dfthalf potential for pristine diamond and the defect hydrogen atom uses an unaltered pseudopotential. The bulk \dfthalf band structure obtained in this matter can be seen in the middle of Fig. \ref{fig:H_T_schema_BS}. The occupied hydrogen level $H_\uparrow$ of this bulk \dfthalf calculation is then shifted down by $E_{gap}(r_c=r_{c,max})-E_{gap}(r_c=0)$ of the left curve of Fig. \ref{fig:H_T_decoupled} such that the difference between the CBM and this level is exactly the maximum gap determined in this curve. The unoccupied level $H_\downarrow$ is shifted up in the same manner such that the gap between this level and the VBM is that of the curve on the right of Fig. \ref{fig:H_T_decoupled}.

The band structure has to be constructed in this way because the decoupled \dfthalf method uses two different pseudopotentials, one for each defect level. This means that there is no single calculation that can provide these eigenvalues and more importantly no wavefunction.

\begin{figure}[h]
    \centering
    \includegraphics[width=0.45\textwidth]{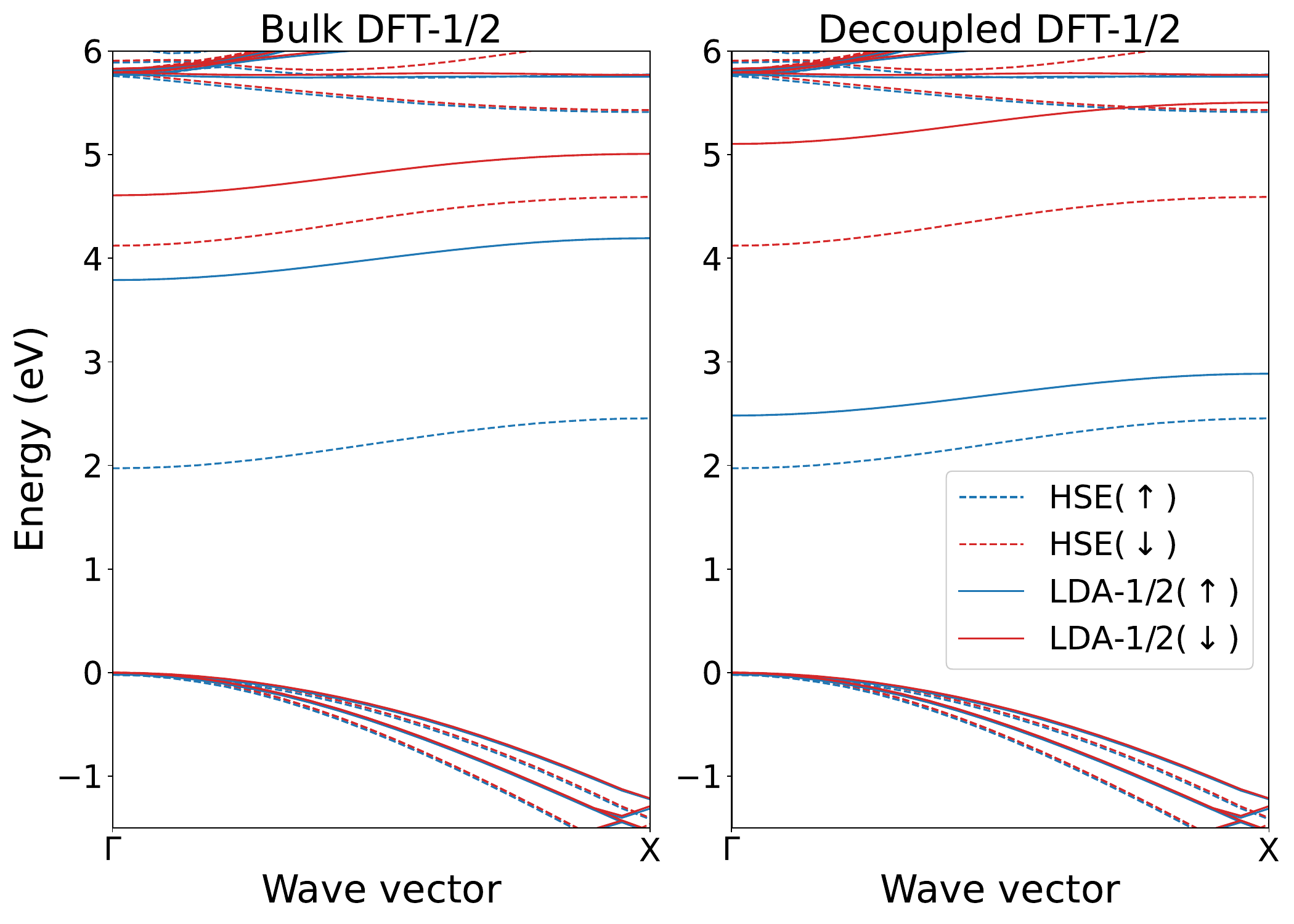}
    \caption{The band structure of tetrahedral hydrogen for a $2\times2\times2$ diamond conventional supercell using bulk \dfthalf and HSE06 (left) and bulk+decoupled, \dfthalf and HSE06 (right). The zero point of energy was set at the VBM for all calculations.}
    \label{fig:H_T_HSE_bs}
\end{figure}

To test whether the decoupled method produces a correct gap between the defect levels, the band structure for the hydrogen interstitial was calculated for a $2\times2\times2$ supercell using the decoupled method and a DFT calculation based on the HSE06 functional \cite{HSE03,HSE06}. 
Because this calculation is meant as a comparison between the electronic structure of the two different methods the same LDA relaxed supercell was used for both calculations. 
In Fig. \ref{fig:H_T_HSE_bs} the HSE06 band structure is compared with both the bulk and bulk + the decoupled \dfthalf method. The band structure of the decoupled \dfthalf method on the right was again constructed by shifting the defect levels, the solid blue and red lines in Fig. \ref{fig:H_T_HSE_bs}, of the bulk \dfthalf band structure by $E_{gap}(r_c=r_{c,max})-E_{gap}(r_c=0)$ such that the gaps between these levels and the VBM and CBM are correct, according the decoupled method. Due to the small size of the supercell, the defect levels show some dispersion for both the \dfthalf and the HSE06 calculations. 
Since the HSE06 and the \dfthalf band gap for diamond differ by about $0.3$ eV this is only a qualitative comparison. For both calculations we placed the zero-point of energy at the VBM. The HSE06 calculation finds a gap between defect levels of about $2.1$ eV, which more closely matches the gap of $2.7$ eV found by the decoupled method, as opposed to the gap found by the bulk \dfthalf method of $0.9$ eV. The position of the defect levels with respect to the VBM and CBM of the decoupled method also matches the HSE06 result the closest. 
This leads us to conclude that if the conventional method fails, the decoupled method results in a qualitatively better result than applying no correction at all.

Although the decoupled method was introduced as a solution for when the conventional method fails to determine the gap between two defect levels, it is demonstrated in Fig. \ref{fig:H_T_HSE_bs} that the entire band structure can be corrected by calculating $E_{VBM \rightarrow unocc}$ and $E_{occ \rightarrow CBM}$ for all unoccupied and occupied defect levels, respectively.

\subsection{The negatively charged vacancy in diamond}

The negatively charged vacancy $V(-)$ consists of 5 electrons, 4 coming from the dangling carbon bonds and 1 from the extra negative charge. Experimentally it has been shown that in the ground state $V(-)$ has the $^4A_2$ many-body state with $T_d$ symmetry \cite{PhysRevB.45.1436}. 
Only the single-particle configuration $a_1(\uparrow,\downarrow)t_2(\uparrow,\uparrow,\uparrow)$ contributes to the $^4A_2$ may body state \cite{PhysRevB.45.1436,GaliDeak_FormationNV,PhysRevB.51.6984}, 
which is stable against Jahn-Teller distortion \cite{PhysRevB.51.6984}. The negatively charged vacancy is also responsible for the GR1 band and the corresponding ZPL at $3.15 $ eV \cite{PhysRevB.46.13157}. This ZPL is associated with the transition from $^4A_2\rightarrow$ $^4T_1$, where the excited $^4T_1$ state can be written in terms of $a_1(\uparrow)t_2(\uparrow \downarrow,\uparrow,\uparrow)$ single particle states \cite{GaliDeak_FormationNV,PhysRevB.51.6984}. Besides the negative charged state the vacancy can also be found in the neutral charge state with a ZPL of $1.67$ eV \cite{PhysRevB.46.13157}, and theoretically it is also shown that the vacancy can be stable with a charge ranging from $-2$ to $2$ \cite{GaliDeak_FormationNV}.

In what follows we try to determine the ZPL of $V(-)$ with the \dfthalf method similar to what Lucatto et al. did for the $NV^{-}$ \cite{dfthalf_NVmin}. We show that the conventional method of Lucatto et al. will not work for this defect and the decoupled method is required. However, the decoupled method also has difficulties with this defect. We provide an explanation for the problems that the decoupled method faces with this defect and offer a solution.

\begin{figure}[h]
    \centering
    \includegraphics[width=0.5\textwidth]{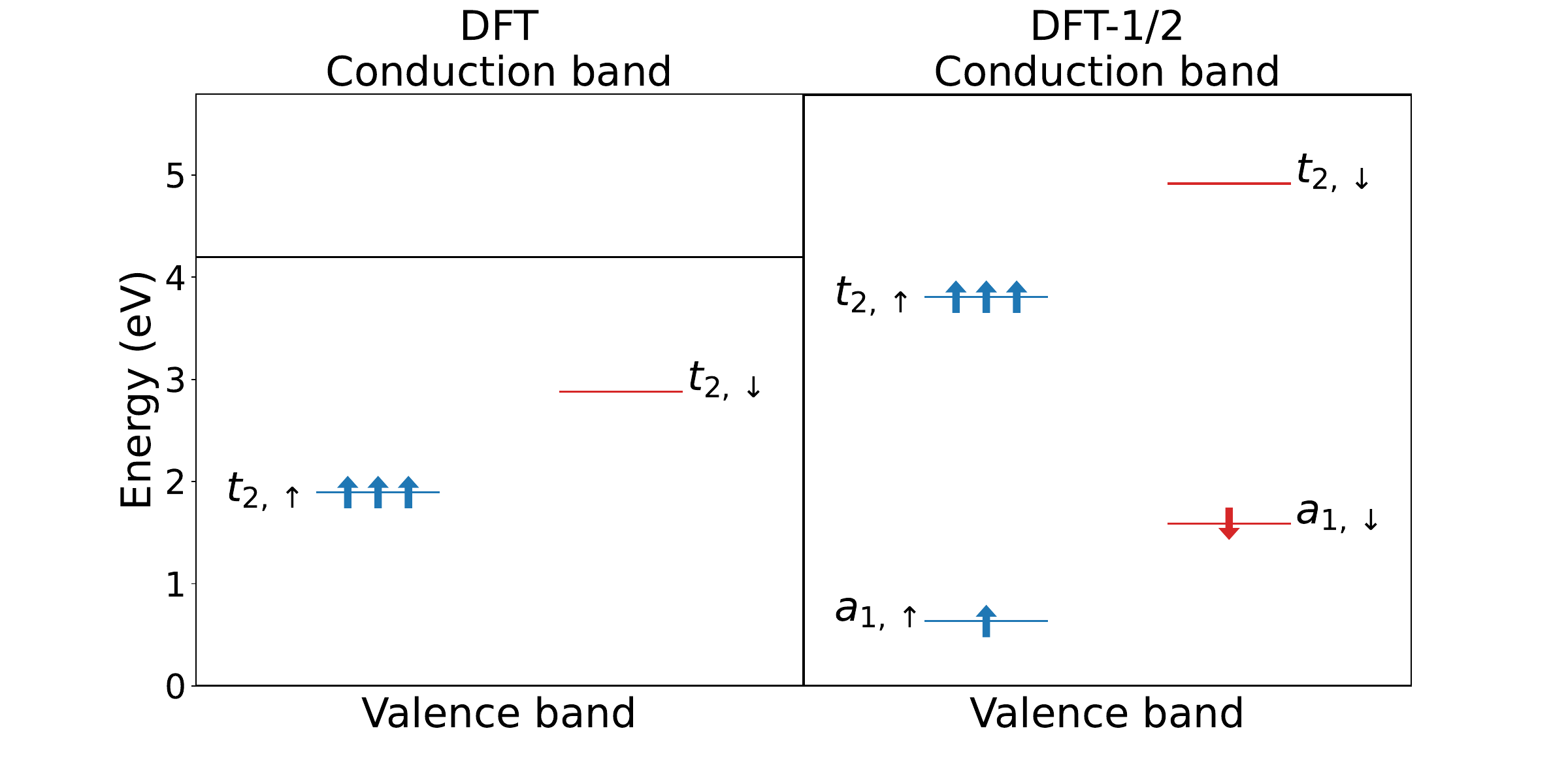}
    \caption{The eigenvalues in the \gammapoint for the negatively charged vacancy in diamond using DFT (left) and bulk DFT$-\frac{1}{2}$ (right), where the defect atoms are the 4 carbon atoms around the vacancy.}
    \label{fig:Vmin_schema_BS_DFT_DFThalf}
\end{figure}

As a first step the electronic structure of $V(-)$ was calculated using DFT and bulk \dfthalf, as depicted in Fig. \ref{fig:Vmin_schema_BS_DFT_DFThalf}. For the bulk \dfthalf calculation the four carbon atoms closest to the vacancy $C_{vac}$ were treated as defect atoms. 
In the \dfthalf band structure both the $a_1$ and triple degenerate $t_2$ levels can be recognized, while the DFT band structure only has the triple degenerate $t_2$ levels. This makes the need for the band gap correction for this defect apparent, because without this correction the $a_1$ levels are hidden in the valence band.

To calculate the ZPL one not only requires the ground state but also the first excited state (see Fig. \ref{fig:schema_ZPL}), which is created by exciting an electron from $a_{1,\downarrow}$ to the $t_{2,\downarrow}$ state. The many-body excited state was approximated by placing one electron in one of the three triple degenerate $t_{2,\downarrow}$ levels using constrained DFT. Then the electron fractions $\xi$ and $\zeta$ for the transition $a_{1,\downarrow} \leftrightarrow t_{2,\downarrow}$ were determined for both the ground and excited state, which can be found in table \ref{tab:Vmin_Efrac_Cvac}. To obtain the defect gap correction only the decoupled method can be used since $\xi_{C,p}\approx\zeta_{C,p}$.

\begin{table}[h]
    \centering
    \begin{tabular}{CCCCC}
    \toprule
    \multicolumn{1}{l}{} &
    \multicolumn{2}{c}{ground state}      &
    \multicolumn{2}{c}{excited state} \\ 
    \cmidrule(lr){2-3}
    \cmidrule(lr){4-5}
         & \xi_{X_\phi}     & \zeta_{X_\phi}    & \xi_{X_\phi}  & \zeta_{X_\phi} \\ \midrule
       C_{vac,s}  & 0.00    & 0.01              & 0.00          & 0.01\\
       C_{vac,p}  & 0.12    & 0.11              & 0.12          & 0.12\\ \bottomrule
    \end{tabular}
    \caption{The electron fraction for the transition $^4A_2\leftrightarrow$ $^4T_1$ of $V(-)$.}
    \label{tab:Vmin_Efrac_Cvac}
\end{table}

\begin{figure}[h]
    \centering
    \includegraphics[width=0.45\textwidth]{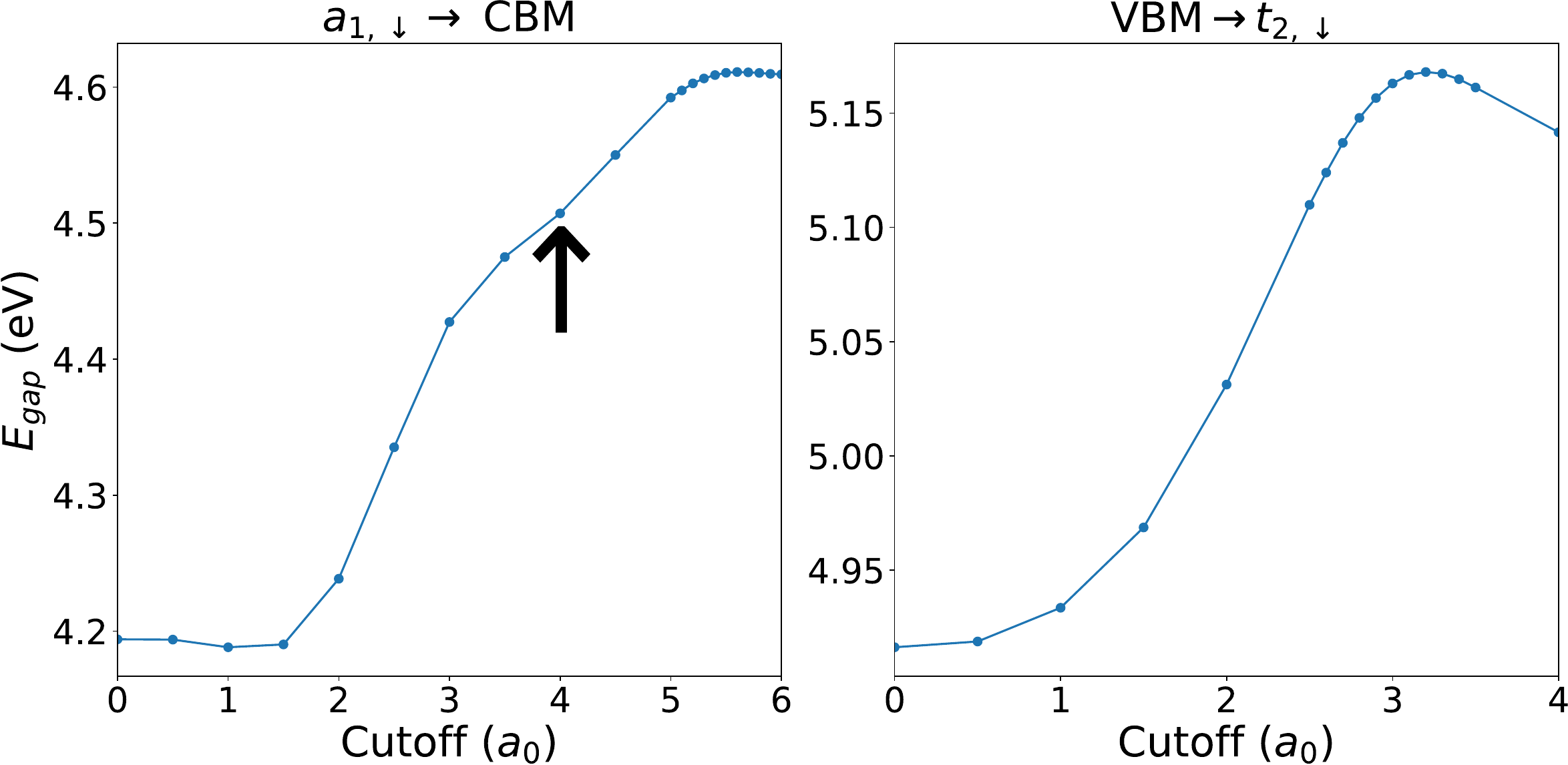}
    \caption{The maximization of the gaps $E_{VBM \rightarrow unocc}$ and $E_{occ \rightarrow CBM}$ for the absorption energy of the $V(-)$ center. An arrow was added to the $E_{occ \rightarrow CBM}$ curve to highlight the kink in the curve.}
    \label{fig:Vmin_Cvac_Eabs_decoupled}
\end{figure}

In what follows we focus on the calculation of the absorption energy with \dfthalf and the problems that come with it. The same story applies to the emission energy. In Fig. \ref{fig:Vmin_Cvac_Eabs_decoupled} the gaps $E_{VBM \rightarrow unocc}$ and $E_{occ \rightarrow CBM}$ are determined for the absorption energy. 
For the gap $E_{occ \rightarrow CBM}$ an unreasonably large cutoff of $5.6 \enspace a_0$ was found, keeping in mind that the bond length between bulk carbon atoms is about $2.9 \enspace a_0$ in our calculations. 
A cutoff this large encompasses the nearest neighbor of the $C_{vac}$, the entire spherical bulk self-energy of the nearest neighbor, the next nearest neighbors and the next next nearest neighbors. The absorption energy and the ZPL based on the maximum gaps of Fig. \ref{fig:Vmin_Cvac_Eabs_decoupled} are $4.0$ eV and $3.9$ eV, respectively. This deviates far from the previously mentioned experimental results. For the gap $E_{VBM \rightarrow unocc}$ a more reasonable cutoff of $3.2 \enspace a_0$ was found.

We suspect that this unreasonably large cutoff is due to the other carbon atoms, that are located close to the vacancy and which also contribute to the occupied defect level. 
The curve of the gaps \emph{occ to CBM} of Fig. \ref{fig:Vmin_Cvac_Eabs_decoupled} seems to have a kink at $r_c=4.0 \enspace a_0$ due to two competing effects working on the defect gap, one with a maximum between $2.5 \enspace a_0$ and $4.0 \enspace a_0$ and one with a maximum around the global maximum of $5.6 \enspace a_0$. 
The first maximum incorporates the nearest neighbors and most of the spatial region of its self-energy since the bulk cutoff is $2.3 \enspace a_0$. The second maximum incorporates the next and next-next nearest neighbors of the $C_{vac}$ atoms. It is as if adding self-energy to these neighboring atoms also increases the gap between the defect levels. It should be noted that the self-energy potential added to the neighbors of the $C_{vac}$ atoms is not the correct self-energy potential for these atoms and these atoms already have a self-energy potential stemming from the bulk \dfthalf correction.  

Upon closer inspection some other carbon atoms located near the vacancy have a nonzero contribution to the defect levels. If these atoms were to be included in the defect atom group they would have a $\xi$ and $\zeta$ of the order of $0.01$. 
The gaps $E_{VBM \rightarrow unocc}$ and $E_{occ \rightarrow CBM}$ where recalculated with the extra defect atoms in an attempt to prevent the large cutoff parameter by adding the correct self-energy potential to these atoms. 
However, this did not improve the results. Therefore, the neighboring atoms might not be the only reason for the large cutoff $r_c$ or the incorrect self-energy was not the cause of the problem.

Since the cutoff parameter for $E_{occ \rightarrow CBM}$ cannot be found by maximizing this gap, the bulk parameter $r_c=2.3 \enspace a_0$ was used instead for both $E_{VBM \rightarrow unocc}$ and $E_{occ \rightarrow CBM}$. We argue that the cutoff parameter should be transferable and will not change much in different chemical environments, as is the case for the cutoff parameter in the bulk \dfthalf method \cite{Ferreira2008}. 
Even in cases where the cutoff parameter can be calculated, taking a small deviation from this value by choosing the bulk cutoff parameter will not influence the gap greatly since close to the maximum the value of the gap is approximately constant. In Ref. \cite{ChargeTransMn} the cutoff parameter is determined for a Mn interstitial and substitutional defect in Si. This was done by following the energy of the defect level in the density of states with respect to the CBM as a function of $r_c$, similar to the \emph{occ to CBM} in this work, resulting in a cutoff of $3.0 \enspace a_0$ for Mn. This cutoff for Mn was also found in Ref. \cite{GaMnAs_pos_Mn_dlevels} where they use \dfthalf on GaMnAs. We further motivate this approach by determining the cutoff parameter using the conventional method. Since $\xi_{C,p}\approx\zeta_{C,p}$ in case of $V(-)$ but not $\xi_{C,p}=\zeta_{C,p}$, the effect on the gap of the \dfthalf approach is severely reduced but the method still produces the correct cutoff parameter of around $2.2 \enspace a_0$ which is close to the bulk cutoff of $2.3 \enspace a_0$. 

\begin{figure}
    \centering
    \includegraphics[width=0.30\textwidth]{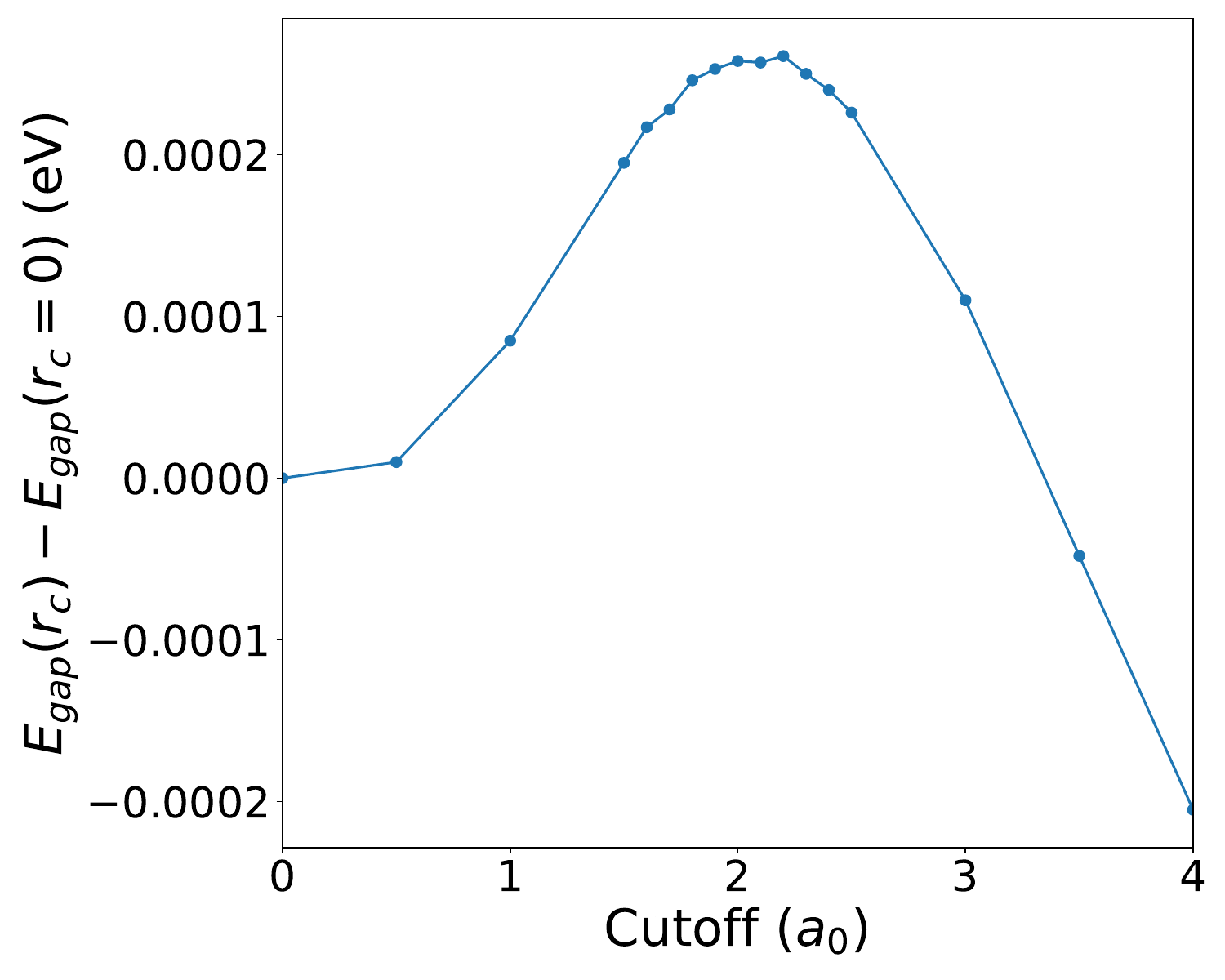}
    \caption{The gap extremization of the absorption energy of $V(-)$ using the conventional \dfthalf method.}
    \label{fig:Vmin_Eabs_conv}
\end{figure}

The gaps $E_{VBM \rightarrow unocc}$ and $E_{occ \rightarrow CBM}$ were then calculated using the bulk cutoff parameter for both the ground and excited state such that the absorption and emission energy could be calculated using formula \eqref{eq:Egap_dec}. With this the ZPL was calculated for the $V(-)$ center. The values of the optical transitions can be found in table \ref{tab:Vmin_ZPL_bulk_cutoff}. In the best case we find a ZPL equal to the one found in \cite{GaliDeak_FormationNV} using HSE06. Our results seem to overestimate the optical transition energies, this is likely due to the approximated cutoff parameter and the error caused by the bulk \dfthalf correction which is about $\pm 0.23$eV in the case of diamond.

\begin{table}
    \centering
    \begin{tabular}{CCCCC}\toprule
                & \text{optical transition} & \text{ZPL} & \text{shift} & \text{(eV)}\\ \midrule
        E_{abs} & 3.64 & 3.47 & 0.16\\
        E_{em}  & 3.08 & 3.30 & 0.22\\ \bottomrule
    \end{tabular}
    \caption{The ZPL of $V(-)$ calculated based on the absorption and emission energy using the bulk cutoff parameter. The last column contains the Stokes and anti-Stokes shift depending on which is required to calculate the ZPL. }
    \label{tab:Vmin_ZPL_bulk_cutoff}
\end{table}

\subsection{The general procedure of the decoupled \dfthalf method}
\noindent We now give a brief overview of the decoupled \dfthalf method.
\subsection*{Preparation \dfthalf}
\begin{enumerate}
    \item Use the \dfthalf method to correct the band gap of the host material.
    \item Make the supercell with the defect and relax it with DFT.
    \item Categorize all atoms in the supercell either as defect or bulk atoms.
    \item Perform a scf calculation where the bulk atoms use the \dfthalf pseudopotential obtained in step 1, while the defect atoms use their unaltered pseudopotential.
    \item Use the orbital contribution of each defect atom to the occupied and unoccupied level from the previous calculation to determine the electron fractions $\xi_{X_\phi}$ and $\zeta_{X_\phi}$.
    \item If there are orbitals for which $\xi_{X_\phi}\approx \zeta_{X_\phi}$ use the decoupled \dfthalf method. Otherwise use the conventional method as described by Lucatto et al. \cite{dfthalf_NVmin}.
\end{enumerate}

\subsection*{The decoupled \dfthalf method}
\noindent For each group of symmetrically equivalent atoms $X$ (or orbitals $X_\phi$) follow the steps below and use the \dfthalf pseudopotential obtained for these atoms while performing the calculations for the next group of atoms.
\begin{enumerate}
    \item Make two sets of \dfthalf pseudopotentials, one where the electron fraction $\xi_{X_\phi}$ is added to $X_\phi$ for every value of the cutoff parameter $r_c$ and one set where the $\zeta_{X_\phi}$ is added.
    \item Determine the optimal cutoff parameter for each set of pseudopotentials by running separate DFT calculations for each value of $r_c$ in each set. The optimal cutoff parameter for $\xi_{X_\phi}$ is the one which extremizes the gap between the occupied defect level and the CBM. The optimal cutoff for $\zeta_{X_\phi}$ should extremize the gap between the VBM and the unoccupied defect level.
    \item Calculate the gap between the defect levels with the two extremes from the previous step and Eq. \ref{eq:Egap_dec}.
    \item Repeat the process for the next defect atom (or orbital).
\end{enumerate}

\section{Conclusions}
To summarize and conclude, the conventional \dfthalf method for defects has been shown to fail in cases where the occupied and unoccupied defect level have a similar orbital character. This is due to the cancellation of the self-energy potential of the unoccupied level by the occupied level. When the electron fractions $\xi_{X,\phi}$ and $\zeta_{X,\phi}$ are large this can be problematic and one should use the decoupled \dfthalf method, which overcomes this problem by decoupling the effect of $\xi_{X,\phi}$ and $\zeta_{X,\phi}$. The decoupled method was tested on a tetrahedral hydrogen interstitial and the negatively charged vacancy in diamond. In the case of the hydrogen interstitial, it was shown that the decoupled \dfthalf method increases the gap between defect levels significantly which was qualitatively more in line with the HSE06 results. This comparison with HSE06 also showed that the decoupled \dfthalf method can be used to correct the entire band structure. 
For the negatively charged vacancy no proper cutoff parameters could be found because the localization of the defect atoms extended further than the nearest neighbors. Instead, the bulk cutoff parameter was used. The addition of this self-energy potential still leads to a ZPL similar to those calculated with HSE06. Although the decoupled method works in cases where the conventional \dfthalf method for defects fails it also has flaws. The decoupled method is sensitive to errors made by the bulk \dfthalf method, it sometimes does not find an appropriate cutoff as was the case with $V(-)$ and it doubles the computational cost. 
\\

 \section*{Acknowledgements}
This work was supported by the FWO (Research Foundation-Flanders), project G0D1721N. 
This work was performed in part using HPC resources from the VSC (Flemish Supercomputer Center) and the HPC infrastructure of the University of Antwerp (CalcUA), both funded by the FWO-Vlaanderen and the Flemish Government department EWI (Economie, Wetenschap \& Innovatie).


\bibliography{references}

\end{document}